\documentclass[preprint,pre,aps,showpacs,amsmath]{revtex4}
\usepackage{graphicx}

\begin{document}

\title{Thermodynamic Order Parameters and Statistical-Mechanical Measures for Characterization of the Burst and Spike Synchronizations
       of Bursting Neurons}

\author{Sang-Yoon Kim}
\email{sangyoonkim@dnue.ac.kr}
\author{Woochang Lim}
\email{woochanglim@dnue.ac.kr}
\affiliation{Computational Neuroscience Lab., Department of Science Education, Daegu National University of Education, Daegu 705-115, Korea}

\begin{abstract}
We are interested in characterization of population synchronization of bursting neurons which exhibit both the slow bursting and the fast spiking timescales, in contrast to spiking neurons. Population synchronization may be well visualized in the raster plot of neural spikes which can be obtained in experiments. The instantaneous population firing rate (IPFR) $R(t)$, which may be directly obtained from the raster plot of spikes, is often used as a realistic collective quantity describing population behaviors in both the computational and the experimental neuroscience.  For the case of spiking neurons, realistic thermodynamic order parameter and statistical-mechanical spiking measure, based on $R(t)$, were introduced in our recent work to make practical characterization of spike synchronization. Here, we separate the slow bursting and the fast spiking timescales via frequency filtering, and extend the thermodynamic order parameter and the statistical-mechanical measure to the case of bursting neurons. Consequently, it is shown in explicit examples that both the order parameters and the statistical-mechanical measures may be effectively used to characterize the burst and spike synchronizations of bursting neurons.
\end{abstract}

%\keywords{Bursting Neurons, Burst and Spike Synchronizations, Thermodynamic Order Parameters and Statistical-Mechanical Measures}

\maketitle

\section{Introduction}
\label{sec:INT}

In recent years, brain rhythms which are observed in scalp electroencephalogram and local field potentials have attracted much attention \cite{Buz}. These brain rhythms emerge via synchronization between individual neuronal firings. Synchronization of firing activities may be used for efficient sensory and cognitive processing (e.g., feature integration, selective attention, and memory formation) \cite{Wang1,Wang2,Gray}. This kind of neural synchronization is also correlated with pathological rhythms associated with neural diseases such as epilepsy, Parkinson's disease, and Alzheimer's disease \cite{Disease1,Disease2,Disease3}. Here, we are interested in characterization of these synchronous brain rhythms.

There are two basic types of neuronal firing activities, spiking and bursting \cite{Izhi1}. We are concerned about synchronization of bursting neurons. Bursting occurs when neuronal activity alternates, on a slow timescale, between a silent phase and an active (bursting) phase of fast repetitive spikings \cite{Burst1,Burst2,Rinzel1,Rinzel2,Burst3}. In neural information transmission, burst input is more likely to have a stronger impact on the postsynaptic neuron than single spike input \cite{Izhi2,Burst4,Burst5,Burst6}. Intrinsically bursting neurons and chattering neurons in the cortex, thalamocortical relay neurons, thalamic reticular neurons, hippocampal pyramidal neurons, Purkinje cells in the cerebellum, pancreatic $\beta$-cells, and respiratory neurons in pre-Botzinger complex are representative examples of bursting neurons \cite{Burst2,Burst3}. These bursting neurons exhibit two different patterns of synchronization due to the slow and fast timescales of bursting activity. Burst synchronization (synchrony on the slow bursting timescale) refers to a temporal coherence between the active phase onset or offset times of bursting neurons, while spike synchronization (synchrony on the fast spike timescale) characterizes a temporal coherence between intraburst spikes fired by bursting neurons in their respective active phases \cite{Burstsync1,Burstsync2}. Recently, many studies on the burst and spike synchronizations have been made in several aspects (e.g., chaotic phase synchronization, transitions between different states of burst synchronization, effect of network topology, effect on information transmission, suppression of bursting synchronization, and effect of noise and coupling on burst and spike synchronizations) \cite{BSsync1,BSsync2,BSsync3,BSsync4,BSsync5,BSsync6,BSsync7,BSsync8,BSsync9,BSsync10,BSsync11,BSsync12,BSsync13,BSsync14,BSsync15}.

In this paper, we are concerned about practical characterization of the burst and spike synchronizations of bursting neurons. Population synchronization may be well visualized in the raster plot of neural spikes which can be obtained in experiments. Instantaneous population firing rate (IPFR), $R(t)$, which is directly obtained from the raster plot of spikes, is a realistic collective quantity describing population behaviors in both the computational and the experimental neuroscience \cite{Wang1,Sparse1,Sparse2,Sparse3,Sparse4,Sparse5,Sparse6}. We note that the experimentally-obtainable $R(t)$ is in contrast to the ensemble-averaged potential $X_G$ which is often used as a population quantity in the computational and theoretical neuroscience, because to directly obtain $X_G$ in real experiments is very difficult. To overcome this difficulty, instead of $X_G$, we employed $R(t)$ as a population quantity, and developed realistic measures, based on $R(t)$, to make practical characterization of synchronization of spiking neurons in both the computational and the experimental neuroscience \cite{Kim}. The mean square deviation of $R(t)$ plays the role of an order parameter $\cal {O}$ used for characterizing synchronization transition of spiking neurons \cite{Order}. The order parameter $\cal {O}$ can be regarded as a ``thermodynamic'' measure because it concerns just the macroscopic quantity $R(t)$ without considering any quantitative relation between $R(t)$ and the microscopic individual spikes. Through calculation of $\cal {O}$, one can determine the threshold value for the spike synchronization. Moreover, to quantitatively measure the degree of spike synchronization, a ``statistical-mechanical'' spiking measure $M_s$ was introduced by taking into consideration both the occupation pattern and the pacing pattern of spikes in the raster plot. Particularly, the pacing degree between spikes was determined in a statistical-mechanical way by quantifying the average contribution of (microscopic) individual spikes to the (macroscopic) IPFR $R(t)$. Consequently, synchronization of spiking neurons may be well characterized in terms of these realistic thermodynamic order parameter and statistical-mechanical measure, $\cal {O}$ and $M_s$, based on $R(t)$.

The main purpose of our work is to characterize the burst and spike synchronizations of bursting neurons by extending the thermodynamic order parameter and the statistical-mechanical measure of spiking neurons \cite{Kim} to the case of bursting neurons. For this aim, we separate the slow and fast timescales of the bursting activity via frequency filtering, and decompose the IPFR $R(t)$ into $R_b(t)$ (the instantaneous population burst rate (IPBR) describing the bursting behavior) and $R_s(t)$ (the instantaneous population spike rate (IPSR) describing the intraburst spiking behavior). Then, the mean square deviations of $R_b$ and $R_s$ play the role of realistic thermodynamic order parameters, ${\cal {O}}_b$ and ${\cal {O}}_s$, used to determine the bursting and spiking thresholds for the burst and spike synchronization, respectively. We also consider another raster plot of bursting onset or offset times for more direct visualization of bursting behavior. From this type of raster plot, we can directly obtain the IPBR, $R_b^{(on)}(t)$ or $R_b^{(off)}(t)$, without frequency filtering. Then, the time-averaged fluctuations of $R_b^{(on)}(t)$ and $R_b^{(off)}(t)$ also play the role of the order parameters, ${\cal {O}}_b^{(on)}$ and ${\cal {O}}_b^{(off)}$, for the bursting transition. These bursting order parameters ${\cal {O}}_b^{(on)}$ and ${\cal {O}}_b^{(off)}$ are more direct ones than ${\cal {O}}_b$ because they may be obtained directly without frequency filtering and they yield the same bursting threshold which is obtained through calculation of ${\cal {O}}_b$. As a next step, in the whole region of burst synchronization, the degree of burst synchronization seen in the raster plot of bursting onset or offset times may be well measured in terms of a statistical-mechanical bursting measure $M_b$, introduced by considering both the occupation and the pacing patterns of bursting onset or offset times in the raster plot. In a similar way, we also develop a statistical-mechanical spiking measure $M_s$, based on $R_s$, to quantitatively measure the degree of the intraburst spike synchronization. Consequently, through separation of the slow bursting and the fast spiking timescales, burst synchronization may be well characterized in terms of both the bursting order parameters (${\cal {O}}_b$, ${\cal {O}}_b^{(on)}$ and ${\cal {O}}_b^{(off)}$) and the statistical-mechanical bursting measure ($M_b$), while characterization of intraburst spike synchronization can be made well by using the spiking order parameter (${\cal {O}}_s$) and the statistical-mechanical spiking measure ($M_s$). To our knowledge, no measures characterizing intraburst spike synchronization of bursting neurons seem to be introduced previously. Hence, ${\cal {O}}_s$ and $M_s$ are new realistic measures characterizing the intraburst spike synchronization. For the case of burst synchronization, our bursting order parameters (${\cal {O}}_b$, ${\cal {O}}_b^{(on)}$ and ${\cal {O}}_b^{(off)}$) and the statistical-mechanical bursting measure ($M_b$) are also in contrast to the conventional measures such as the normalized order parameter $\chi$ \cite{Order1,Order2,Order3,Order4} and the burst phase order parameter $r$ \cite{BSsync1,BSsync4,BSsync6,BPOrder4,BPOrder5}. The normalized order parameter $\chi$ is given through dividing the order parameter (i.e., the time-averaged fluctuation of the ensemble-averaged potential $X_G$) by the average of time-averaged fluctuations of individual potentials. Our bursting order parameters, based on experimentally-obtainable IPBRs, are realistic ones when compared to $\chi$, based on $X_G$, because $X_G$ is very difficult to obtain in experiments. Furthermore, since $X_G$ shows both the bursting and spiking activities, $\chi$ plays the role of an order parameter for the ``whole'' synchronization (including both the burst and spike synchronizations) of bursting neurons, which is in contrast to our bursting order parameters characterizing just the burst synchronization. On the other hand, the burst phase order parameter $r$ is a ``microscopic'' measure quantifying the degree of coherence between  (microscopic) individual burst phases without any explicit relation to the macroscopic occupation and pacing patterns of bursting onset or offset times visualized well in the raster plot, in contrast to our statistical-mechanical bursting measure $M_b$. For our case of $M_b$, the pacing degree (between the bursting onset or offset times) is determined in a statistical-mechanical way by taking into consideration the average of contributions of microscopic individual bursts to the macroscopic IPBR.

This paper is organized as follows. In Sec.~\ref{sec:HR}, as an example for characterization we describe an inhibitory population of bursting Hindmarsh-Rose (HR) neurons \cite{HR1,HR2,HR3,HR6}. In Sec.~\ref{sec:Measure}, through separation of the slow and fast timescales, we develop realistic order parameters and statistical-mechanical measures, based on IPBR and IPSR, which are applicable in both the computational and experimental neuroscience. Their usefulness for characterization of the burst and spike synchronizations is shown in explicit examples of bursting HR neurons. Finally, a summary is given in Section \ref{sec:SUM}.

\section{Inhibitory Population of Bursting Hindmarsh-Rose Neurons}
\label{sec:HR}
As an example for characterization, we consider an inhibitory population of $N$ globally-coupled bursting neurons. As an element in our coupled neural system, we choose the representative bursting HR neuron model which was originally introduced to describe the time evolution of the membrane potential for the pond snails
\cite{HR1,HR2,HR3,HR6}. The population dynamics in this neural network is governed by the following set of ordinary differential equations:
\begin{eqnarray}
\frac{dx_i}{dt} &=& y_i - a x^{3}_{i} + b x^{2}_{i} - z_i +I_{DC} +D \xi_{i} -I_{syn,i}, \label{eq:CHRA} \\
\frac{dy_i}{dt} &=& c - d x^{2}_{i} - y_i, \label{eq:CHRB} \\
\frac{dz_i}{dt} &=& r \left[ s (x_i - x_o) - z_i \right], \label{eq:CHRC} \\
\frac{dg_i}{dt}&=& \alpha g_{\infty}(x_i) (1-g_i) - \beta g_i, \;\;\; i=1, \cdots, N, \label{eq:CHRD}
\end{eqnarray}
where
\begin{eqnarray}
I_{syn,i} &=& \frac{J}{N-1} \sum_{j(\ne i)}^N g_j(t) (x_i - X_{syn}), \label{eq:CHRE} \\
g_{\infty} (x_i) &=& 1/[1+e^{-(x_i-x^*_s)\delta}]. \label{eq:CHRF}
\end{eqnarray}
Here, the state of the $i$th neuron at a time $t$ (measured in units of milliseconds) is characterized by four state variables: the fast membrane potential $x_i$, the fast recovery current $y_i,$ the slow adaptation current $z_i$, and the synaptic gate variable $g_i$ denoting the fraction of open synaptic ion channels. The parameters in the single HR neuron are taken as $a=1.0,$ $b=3.0,$ $c=1.0,$ $d=5.0,$ $r=0.001,$ $s=4.0,$  and $x_o=-1.6$ \cite{HR4}.

\begin{figure}
\includegraphics[width=\columnwidth]{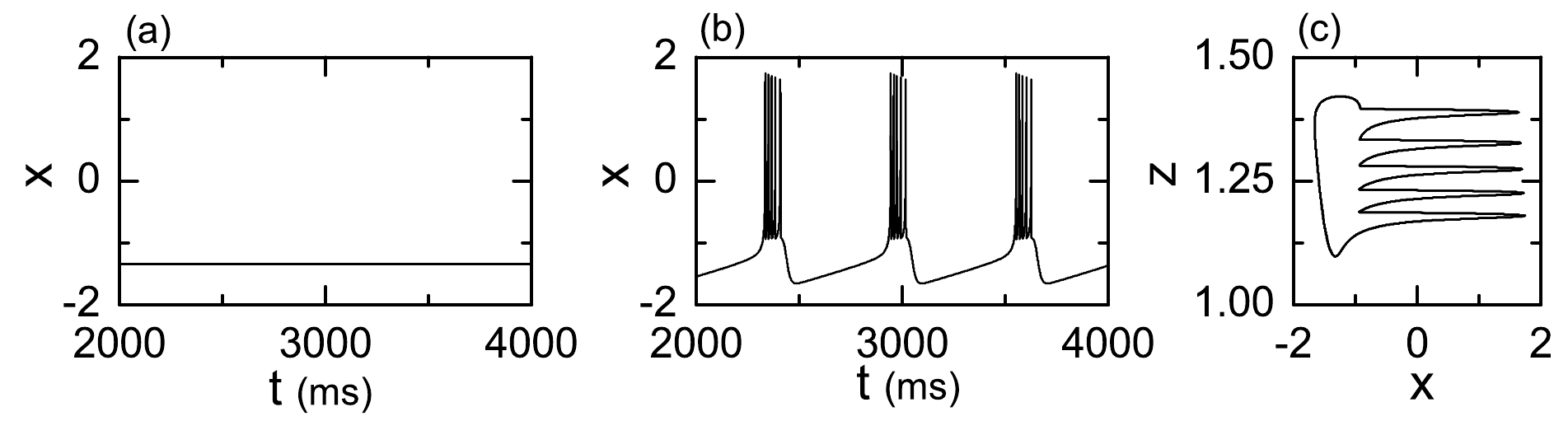}
\caption{Single bursting HR neuron.
Time series of the fast membrane potential $x$ for (a) $I_{DC}=1.2$ and (b) $I_{DC}=1.3$.
(c) Projection of the phase flow onto the $x-z$ plane for $I_{DC}=1.3$.
}
\label{fig:Single}
\end{figure}

Each bursting HR neuron is stimulated by using the common DC current $I_{DC}$ and an independent Gaussian white noise $\xi_i$ [see the 5th and the 6th terms in Eq.~(\ref{eq:CHRA})] satisfying $\langle \xi_i(t) \rangle =0$ and $\langle \xi_i(t)~\xi_j(t') \rangle = \delta_{ij}~\delta(t-t')$, where
$\langle\cdots\rangle$ denotes the ensemble average. The noise $\xi$ is a parametric one that randomly perturbs the strength of the applied current
$I_{DC}$, and its intensity is controlled by using the parameter $D$. As $I_{DC}$ passes a threshold $I_{DC}^* (\simeq 1.26)$ in the absence of noise, each single HR neuron exhibits a transition from a resting state [Fig.~\ref{fig:Single}(a)] to a bursting state [Fig.~\ref{fig:Single}(b)]. As shown in Fig.~\ref{fig:Single}(c), projection of the phase flow onto the $x-z$ plane seems to be a hedgehog-like attractor. Bursting activity [alternating between a silent phase and an active (bursting) phase of fast repetitive spikings] occurs on the hedgehog-like attractor [the body (spines) of the hedgehog-like attractor corresponds to the silent (active) phase]. Here, we consider the suprathreshold case of $I_{DC}=1.3$ where each HR neuron exhibits spontaneous bursting activity without noise.

The last term in Eq.~(\ref{eq:CHRA}) represents the synaptic coupling of the network. $I_{syn,i}$ of Eq.~(\ref{eq:CHRE}) represents a synaptic current injected into the $i$th neuron. Here the coupling strength is controlled by the parameter $J$ and $X_{syn}$ is the synaptic reversal potential. Here, we use $X_{syn}=-2$ for the inhibitory synapse. The synaptic gate variable $g$ obeys the 1st order kinetics of Eq.~(\ref{eq:CHRD}) \cite{Order3,KI2}. Here, the normalized concentration of synaptic transmitters, activating the synapse, is assumed to be an instantaneous sigmoidal function of the membrane potential with a threshold $x^*_s$ in Eq.~(\ref{eq:CHRF}), where we set $x^*_s=0$ and $\delta=30$ \cite{HR5}.
The transmitter release occurs only when the neuron emits a spike (i.e., its potential $x$ is larger than $x^*_s$). For the inhibitory GABAergic synapse (involving the $\rm{GABA_A}$ receptors), the synaptic channel opening rate, corresponding to the inverse of the synaptic rise time $\tau_r$, is $\alpha=10$ ${\rm ms}^{-1}$, and the synaptic closing rate $\beta$, which is the inverse of the synaptic decay time $\tau_d$, is $\beta=0.1$ ${\rm ms}^{-1}$ \citep{GABA1,GABA2}. Hence, $I_{syn}$ rises fast and decays slowly.

Numerical integration of Eqs.~(\ref{eq:CHRA})-(\ref{eq:CHRD}) is done using the Heun method \cite{SDE} (with the time step $\Delta t=0.01$ ms).
For each realization of the stochastic process, we choose a random initial point $[x_i(0),y_i(0),z_i(0),g_i(0)]$ for the $i$th $(i=1,\dots, N)$ neuron with uniform probability in the range of $x_i(0) \in (-2,2)$, $y_i(0) \in (-16,0)$, $z_i(0) \in (1.1,1.4)$, and $g_i(0) \in (0,1)$.

\section{Characterization of The Burst and Spike Synchronizations in Terms of Thermodynamic Order Parameters and Statistical-Mechanical Measures}
\label{sec:Measure}
In this section, we extend the order parameter and the statistical-mechanical measure to the case of bursting neurons for characterization of population synchronization of bursting neurons. For this aim, we separate the slow and fast timescales of the bursting activity via frequency filtering, and decompose the IPFR $R(t)$ into the IPBR $R_b(t)$ (describing the bursting behavior) and the IPSR $R_s(t)$ (describing the intraburst spiking behavior). Then, we develop realistic thermodynamic order parameters and statistical-mechanical measures, based on $R_b(t)$ and $R_s(t)$, and show their usefulness for characterization of the burst and spike synchronizations in explicit examples of HR bursting neurons.

\begin{figure}
\includegraphics[width=0.7\columnwidth]{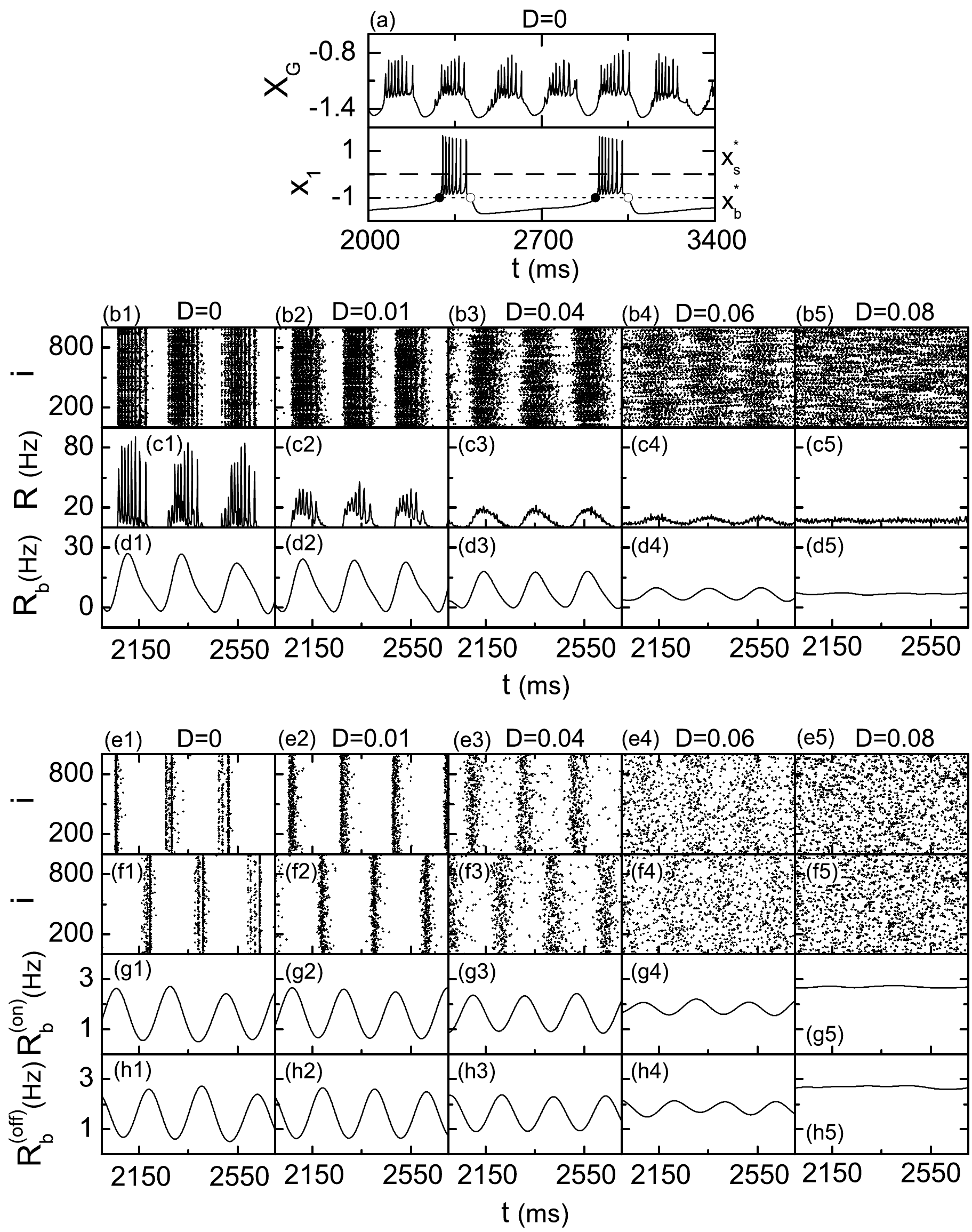}
\caption{Population bursting states for various values of $D$ in an inhibitory ensemble of $N$ $(=10^3)$ globally-coupled bursting HR neurons for $I_{DC}=1.3$ and $J=0.3$: synchronized bursting states for $D=0,$ 0.01, 0.04, and 0.06, and unsynchronized bursting state for $D=0.08$. (a) Time series of the ensemble-averaged global potential $X_G$ and time series of the individual potential $x_1$ of the 1st neuron for $D=0$. The dotted horizontal line ($x^*_b=-1$) represents the bursting threshold (the solid and open circles denote the bursting onset and offset times, respectively), while the dashed horizontal line ($x^*_s=0$) represents the spiking threshold within the active phase. Raster plots of neural spikes for (b1)-(b5), time series of IPFR kernel estimate $R(t)$ for (c1)-(c5), time series of low-pass filtered IPBR $R_b(t)$ (cut-off frequency=10 Hz) for (d1)-(d5), raster plots of active phase (bursting) onset times for (e1)-(e5), raster plots of active phase (bursting) offset times for (f1)-(f5), time series of IPBR kernel estimate $R_b^{(on)}(t)$ for (g1)-(g5), and time series of IPBR kernel estimate $R_b^{(off)}(t)$ for (h1)-(h5). The band width $h$ of the Gaussian kernel function is 1 ms for the IPFR kernel estimate $R(t)$ and 50 ms for the IPBR kernel estimates $R_b^{(on)}(t)$ and $R_b^{(off)}(t)$.
}
\label{fig:Bursting}
\end{figure}

\begin{figure}
\includegraphics[width=0.8\columnwidth]{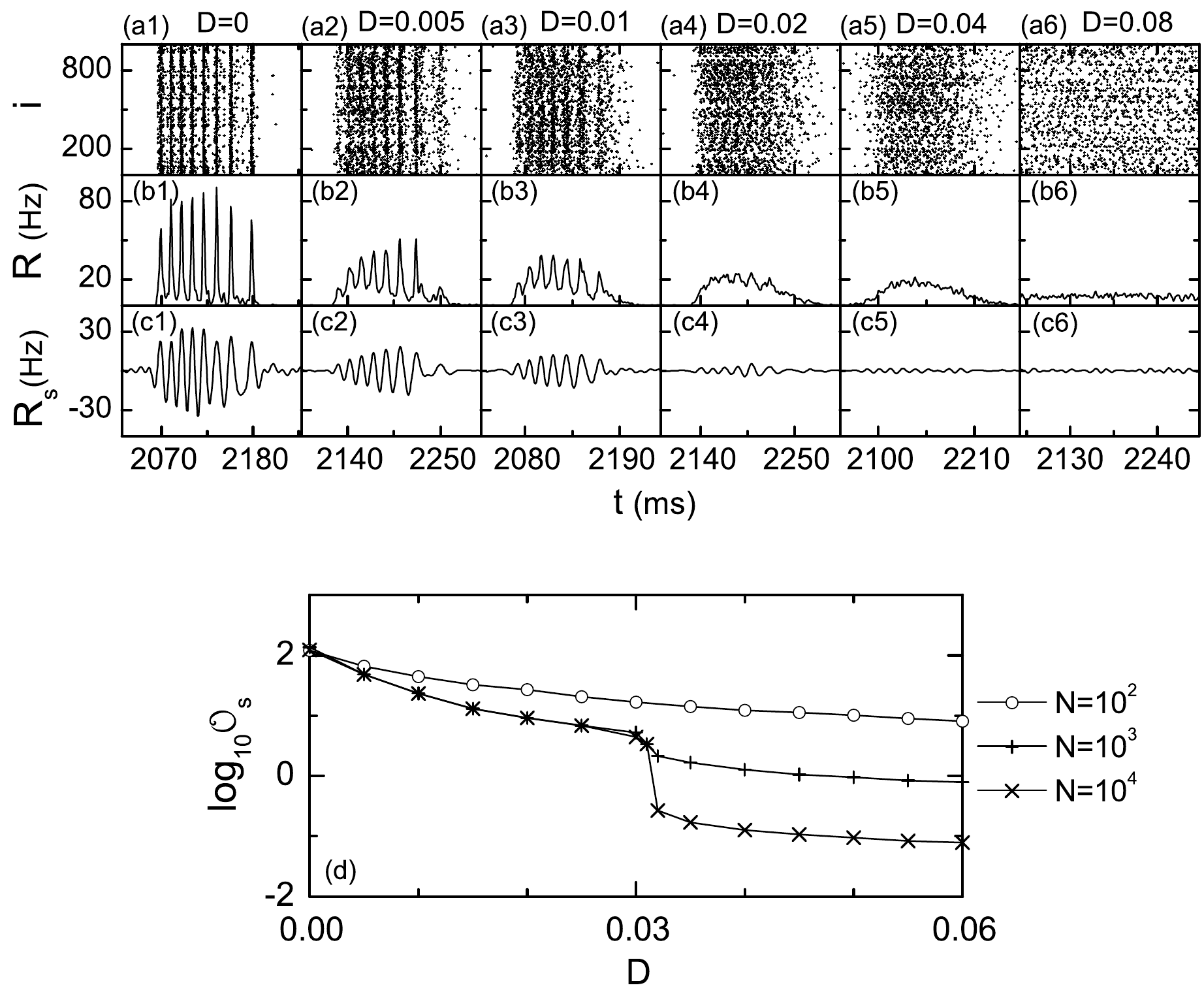}
\caption{Population intraburst spiking states for various values of $D$ and determination of the spiking noise threshold $D^*_s$ for the intraburst spiking transition in terms of the realistic spiking order parameter in an inhibitory ensemble of $N$ globally-coupled bursting HR neurons for $I_{DC}=1.3$ and $J=0.3$: synchronized spiking states for $D=0,$ 0.005, 0.01, and 0.02, and unsynchronized spiking states for $D=0.04$ and 0.08. $N=10^3$ except for the case of (d). (a1)-(a6) Raster plots of neural spikes, (b1)-(b6) time series of IPFR kernel estimate $R(t)$, and (c1)-(c6) time series of band-pass filtered IPSR  $R_s(t)$ [lower and higher cut-off frequencies of 30 Hz (high-pass filter) and 90 Hz (low-pass filter)] in the 1st global bursting cycle of the low-pass filtered IPBR $R_b(t)$ shown in Figs.~\ref{fig:Bursting}(d1)-\ref{fig:Bursting}(d5) (after the transient time of $2 \times 10^3$ ms) for each $D$. Determination of $D^*_s$ for the intraburst spiking transition: (d) plots of spiking order parameters ${\cal{O}}_s$ [based on $R_s(t)$] versus $D$.
}
\label{fig:Spiking}
\end{figure}

As an example for characterization, we consider an inhibitory population of $N$ globally-coupled bursting HR neurons for $I_{DC}=1.3$ and set the coupling strength as $J=0.3$. By varying the noise intensity $D$, we characterize the population synchronization of bursting HR neurons. In computational neuroscience, an ensemble-averaged global potential,
\begin{equation}
 X_G (t) = \frac {1} {N} \sum_{i=1}^{N} x_i(t),
\label{eq:GPOT}
\end{equation}
is often used for describing emergence of population synchronization. Throughout this study, we consider the population behaviors after the transient time of $2 \times 10^3$ ms. Figure \ref{fig:Bursting}(a) shows an oscillating global potential $X_G$ for a synchronous case of  $D=0$. For comparison, an individual potential $x_1$ of the 1st HR neuron is also shown in Fig.~\ref{fig:Bursting}(a). In contrast to $X_G$, each HR neuron fires sparse burstings about once per three global cycles of $X_G$. An active phase of the bursting activity begins at a bursting onset time and ends at a bursting offset time. At the bursting onset (offset) time, each individual potential $x_i$ of the $i$th bursting neuron passes the threshold of $x^*_b = -1$ from below (above); the bursting onset (offset) times of the 1st neuron (after the transient time) are denoted by the solid (open) circles in Fig.~\ref{fig:Bursting}(a). Although the global potential $X_G$ is an important population-averaged quantity to describe synchronization in computational neuroscience, it is practically difficult to directly obtain $X_G$ in real experiments. To overcome this difficulty, instead of $X_G$, we use the IPFR which is an experimentally-obtainable population quantity used in both the experimental and the computational neuroscience \cite{Wang1,Sparse1,Sparse2,Sparse3,Sparse4,Sparse5,Sparse6}. The IPFR is obtained from the raster plot of neural spikes which is a collection of spike trains of individual neurons. Such raster plots of spikes, where population spike synchronization may be well visualized, are fundamental data in experimental neuroscience. As examples of population bursting states, Figs.~\ref{fig:Bursting}(b1)-\ref{fig:Bursting}(b5) show the raster plots of neural spikes for various values of noise intensity $D$: synchronized bursting states for $D=0,$ 0.01, 0.04, and 0.06, and unsynchronized bursting state for $D=0.08$. To obtain a smooth IPFR from the raster plot of spikes, we employ the kernel density estimation (kernel smoother) \cite{Kernel}. Each spike in the raster plot is convoluted (or blurred) with a kernel function $K_h(t)$ to obtain a smooth estimate of IPFR, $R(t)$:
\begin{equation}
R(t) = \frac{1}{N} \sum_{i=1}^{N} \sum_{s=1}^{n_i} K_h (t-t_{s}^{(i)}),
\label{eq:IPSRK}
\end{equation}
where $t_{s}^{(i)}$ is the $s$th spiking time of the $i$th neuron, $n_i$ is the total number of spikes for the $i$th neuron, and we use a Gaussian
kernel function of band width $h$:
\begin{equation}
K_h (t) = \frac{1}{\sqrt{2\pi}h} e^{-t^2 / 2h^2}, ~~~~ -\infty < t < \infty.
\label{eq:Gaussian}
\end{equation}
Figures \ref{fig:Bursting}(c1)-\ref{fig:Bursting}(c5) show smooth IPFR kernel estimates $R(t)$ of band width $h=1$ ms.
For $D=0$, clear ``bursting bands,'' each of which is composed of ``stripes'' of spikes, appear successively at nearly regular time intervals [see Fig.~\ref{fig:Bursting}(b1)]; a magnification of the 1st bursting band is given in Fig.~\ref{fig:Spiking}(a1). For this case of $D=0$, in addition to burst synchronization [synchrony on the slow bursting timescale $\tau_b$ ($\simeq 215$ ms)], spike synchronization [synchrony on the fast spike timescale $\tau_s$ $(\simeq 14.6$ ms)] occurs in each bursting band. As a result of this complete synchronization, the IPFR kernel estimate $R(t)$ exhibits a bursting activity [i.e., fast spikes appear on a slow wave in $R(t)$], as shown in Fig.~\ref{fig:Bursting}(c1). However, as $D$ is increased, loss of spike synchronization begins to occur in each bursting band due to a smearing of spiking stripes. As an example, see the case of $D=0.01$ where the raster plot of spikes and the IPFR kernel estimate $R(t)$ are given in Figs.~\ref{fig:Bursting}(b2) and \ref{fig:Bursting}(c2), respectively. Smearing of the spiking stripes is well seen in the magnified 1st bursting band of Fig.~\ref{fig:Spiking}(a3). Hence, the amplitude of $R(t)$ decreases, as shown in Fig.~\ref{fig:Bursting}(c2). As $D$ is further increased and passes a spiking noise threshold $D^*_s$ $(\simeq 0.032)$, complete loss of spike synchronization occurs in each bursting band (i.e., spikes become incoherent within each bursting band). Consequently, only the burst synchronization (without spike synchronization) occurs [e.g., see the case of $D=0.04$ in Figs.~\ref{fig:Bursting}(b3) and \ref{fig:Bursting}(c3)]. For this case, $R(t)$ shows a slow-wave oscillation without spikes. With increasing $D$, such ``incoherent'' bursting bands become more and more smeared, and thus the degree of burst synchronization decreases [e.g., see Figs.~\ref{fig:Bursting}(b4) and \ref{fig:Bursting}(c4) for $D=0.06$]. Consequently, the amplitude of $R(t)$ is further decreased. With further increase in $D$, incoherent bursting bands begin to overlap, which eventually results in the complete loss of burst synchronization when passing a larger bursting noise threshold $D^*_b$ $(\simeq 0.068)$. In this way, completely unsynchronized states with nearly stationary $R(t)$ appear, as shown in Figs.~\ref{fig:Bursting}(b5) and \ref{fig:Bursting}(c5) for $D=0.08$.

We note that the IPFR kernel estimate $R(t)$ is a population quantity describing the ``whole'' combined collective behaviors (including both the burst and spike synchronizations) of bursting neurons. For more clear investigation of population synchronization, we separate the slow bursting timescale and the fast spiking timescale via frequency filtering, and decompose the IPFR kernel estimate $R(t)$ into the IPBR $R_b(t)$ and the IPSR $R_s(t)$. Through low-pass filtering of $R(t)$ with cut-off frequency of 10 Hz, we obtain the regularly-oscillating IPBR $R_b(t)$ (containing only the bursting behavior without spiking) in Figs.~\ref{fig:Bursting}(d1)-\ref{fig:Bursting}(d5). As $D$ is increased, the amplitude of $R_b(t)$ decreases gradually, and eventually $R_b(t)$ becomes nearly stationary when passing the bursting noise threshold $D^*_b$. For more direct visualization of bursting behavior, we consider another raster plot of bursting onset or offset times, from which we can directly obtain the IPBR kernel estimate of band width $h=50$ ms, $R_b^{(on)}(t)$ or $R_b^{(off)}(t)$, without frequency filtering. Figures \ref{fig:Bursting}(e1)-\ref{fig:Bursting}(e5) show the raster plots of the bursting onset times, while the raster plots of the bursting offset times are shown in Figs.~\ref{fig:Bursting}(f1)-\ref{fig:Bursting}(f5). From these raster plots of the bursting onset (offset) times, we obtain smooth IPBR kernel estimates, $R_b^{(on)}(t)$ [$R_b^{(off)}(t)$] in Figs.~\ref{fig:Bursting}(g1)[(h1)]-\ref{fig:Bursting}(g5)[(h5)]. For $D=0$, clear ``bursting stripes'' [composed of bursting onset (offset) times and indicating burst synchronization] are formed in these raster plots; the bursting onset and offset stripes are time-shifted [see Figs.~\ref{fig:Bursting}(e1) and \ref{fig:Bursting}(f1)]. The corresponding IPBR kernel estimates, $R_b^{(on)}(t)$ and $R_b^{(off)}(t)$, for $D=0$ show regular oscillations with the same population bursting frequency $f_b$ $(\simeq 4.7$ Hz), as shown in Figs.~\ref{fig:Bursting}(g1) and \ref{fig:Bursting}(h1), although they are phase-shifted. As $D$ is increased, the bursting stripes in the raster plots become smeared and begin to overlap, and thus the degree of the burst synchronization decreases. Consequently, the amplitudes of both $R_b^{(on)}(t)$ and $R_b^{(off)}(t)$ decrease gradually (e.g., see the cases of $D=0.01$, 0.04, and 0.06). Eventually, when passing the bursting noise threshold $D^*_b$, bursting onset and offset times become completely scattered in the raster plots, and the corresponding IPBR kernel estimates, $R_b^{(on)}(t)$ and $R_b^{(off)}(t)$, become nearly stationary, as shown in Figs.~\ref{fig:Bursting}(g5) and \ref{fig:Bursting}(h5) for $D=0.08$. In this way, $R_b^{(on)}(t)$ and $R_b^{(off)}(t)$ are more direct ones for describing the bursting behaviors than $R_b(t)$.

\begin{figure}
\includegraphics[width=0.7\columnwidth]{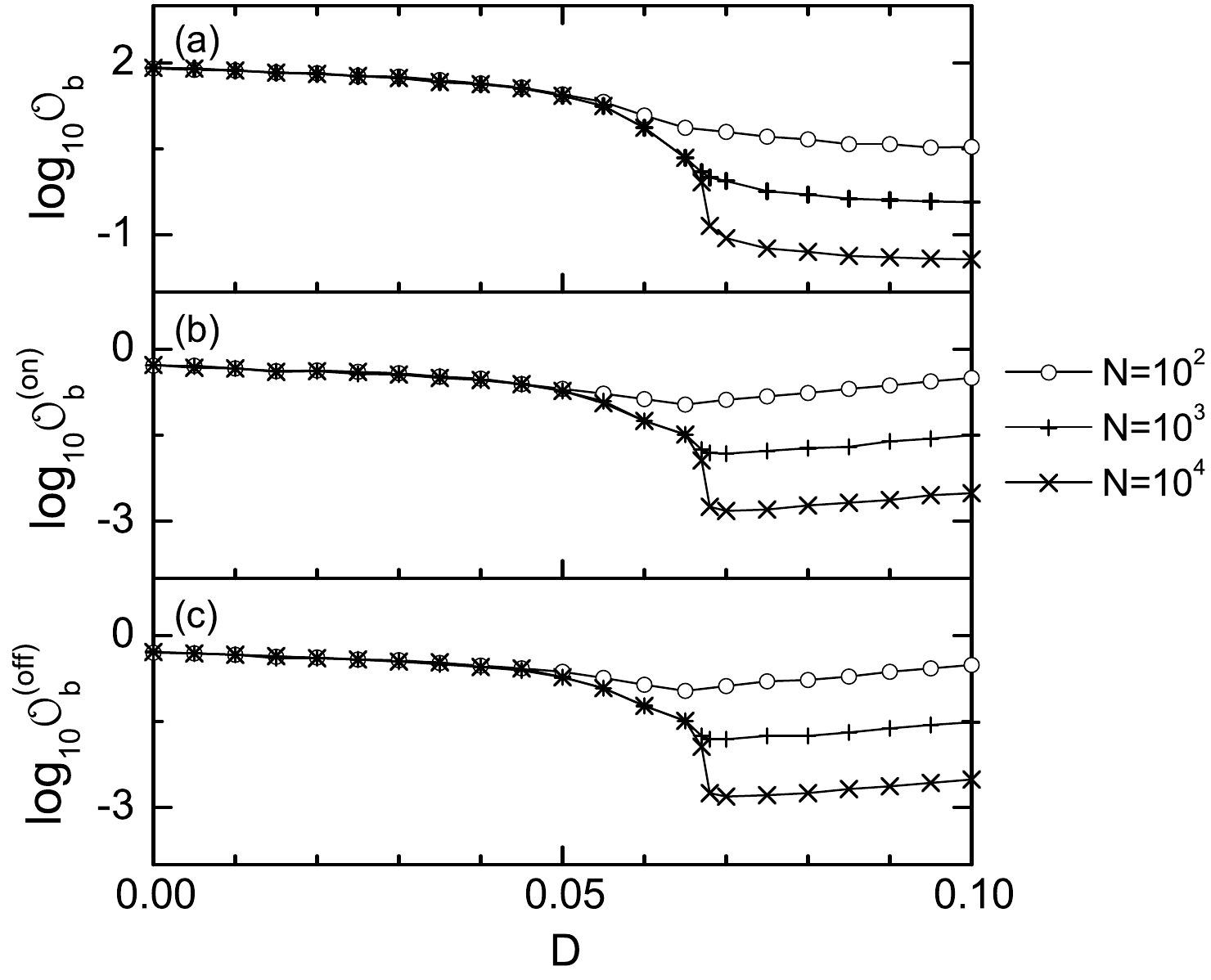}
\caption{Determination of the bursting noise threshold $D^*_b$ for the bursting transition in terms of realistic thermodynamic bursting order parameters in an inhibitory ensemble of $N$ globally-coupled bursting HR neurons for $I_{DC}=1.3$ and $J=0.3$. Plots of bursting order parameters (a) ${\cal{O}}_b$ [based on $R_b(t)$], (b) ${\cal{O}}_b^{(on)}$ [based on $R_b^{(on)}(t)$], and (c) ${\cal{O}}_b^{(off)}$ [based on $R_b^{(off)}(t)$] versus $D$. For each $D$, we compute the bursting order parameters, ${\cal{O}}_b$, ${\cal{O}}_b^{(on)}$, and ${\cal{O}}_b^{(off)}$ by following a trajectory for $3 \times 10^4$ ms after discard the transients for $2 \times 10^3$ ms.
}
\label{fig:Border}
\end{figure}

As is well known, a conventional order parameter, based on the ensemble-averaged global potential $X_G$, is often used for describing transition from asynchrony to synchrony in computational neuroscience \cite{Order1,Order2,Order3,Order4}. Here, instead of $X_G$ which is very difficult to obtain in experiments, we use an experimentally-obtainable IPBR  $R_b(t)$ (which is obtained from the IPFR kernel estimate $R(t)$ via low-pass filtering), and develop a realistic bursting order parameter for the bursting transition, which may be applicable in both the computational and the experimental neuroscience. The mean square deviation of $R_b(t)$,
\begin{equation}
{\cal{O}}_b \equiv \overline{(R_b(t) - \overline{R_b(t)})^2},
 \label{eq:Border1}
\end{equation}
plays the role of a bursting order parameter ${\cal{O}}_b$, where the overbar represents the time average. The order parameter ${\cal{O}}_b$ may be regarded as a thermodynamic measure because it concerns just the macroscopic IPBR $R_b(t)$ without any consideration between $R_b(t)$ and microscopic individual burstings. Here, we discard the first time steps of a trajectory as transients for $2 \times 10^3$ ms, and then we compute ${\cal{O}}_b$ by following the trajectory for $3 \times 10^4$ ms. As $N$ is increased, $R_b(t)$ exhibits more regular oscillations
for the case of burst synchronization, while $R_b(t)$ becomes more stationary for the case of burst unsynchronization. Hence, the bursting order parameter ${\cal{O}}_b$, representing time-averaged fluctuations of $R_b(t)$, approaches a non-zero (zero) limit value for the synchronized (unsynchronized) bursting state in the thermodynamic limit of $N \rightarrow \infty$.
Figure \ref{fig:Border}(a) shows plots of the order parameter ${\cal{O}}_b$ versus $D$. For $D < D^*_b$ $(\simeq 0.068$), synchronized bursting states exist because the values of ${\cal {O}}_b$ become saturated to non-zero limit values. As $D$ passes the bursting noise threshold $D^*_b$, the bursting order parameter ${\cal{O}}_b$ tends to zero as $N \rightarrow \infty$, and hence a transition to unsynchronized bursting states occurs because the noise spoils the burst synchronization. In addition to $R_b(t)$, we also employ another IPBR kernel estimates, $R_b^{(on)}(t)$ and $R_b^{(off)}(t)$, (which are directly obtained from the raster plots of bursting onset and offset times), and introduce realistic bursting order parameters, ${\cal {O}}_b^{(on)}$ and ${\cal {O}}_b^{(off)}$:
\begin{equation}
{\cal{O}}_b^{(on)} \equiv \overline{(R_b^{(on)}(t) - \overline{R_b^{(on)}(t)})^2}~{\rm {and}}~
{\cal{O}}_b^{(off)} \equiv \overline{(R_b^{(off)}(t) - \overline{R_b^{(off)}(t)})^2}.
 \label{eq:Border2}
\end{equation}
Figures \ref{fig:Border}(b) and \ref{fig:Border}(c) show plots of the bursting order parameters ${\cal {O}}_b^{(on)}$ and ${\cal {O}}_b^{(off)}$ versus $D$, respectively. Like the case of ${\cal{O}}_b$, when passing the same bursting noise threshold $D^*_b$, the bursting order parameters
${\cal {O}}_b^{(on)}$ and ${\cal {O}}_b^{(off)}$ also go to zero as $N \rightarrow \infty$, and hence a transition to burst unsynchronization occurs.
In this way, the noise threshold $D^*_b$ for the bursting transition may be well determined through calculation of the three realistic bursting order parameters, ${\cal{O}}_b$, ${\cal {O}}_b^{(on)}$ and ${\cal {O}}_b^{(off)}$. Particularly, ${\cal {O}}_b^{(on)}$ and ${\cal {O}}_b^{(off)}$ are more direct ones than ${\cal {O}}_b$ because they are based on the IPBRs $R_b^{(on)}(t)$ and $R_b^{(off)}(t)$ which are directly obtained from the raster plots of the bursting onset and offset times without frequency filtering, respectively.

\begin{figure}
\includegraphics[width=0.8\columnwidth]{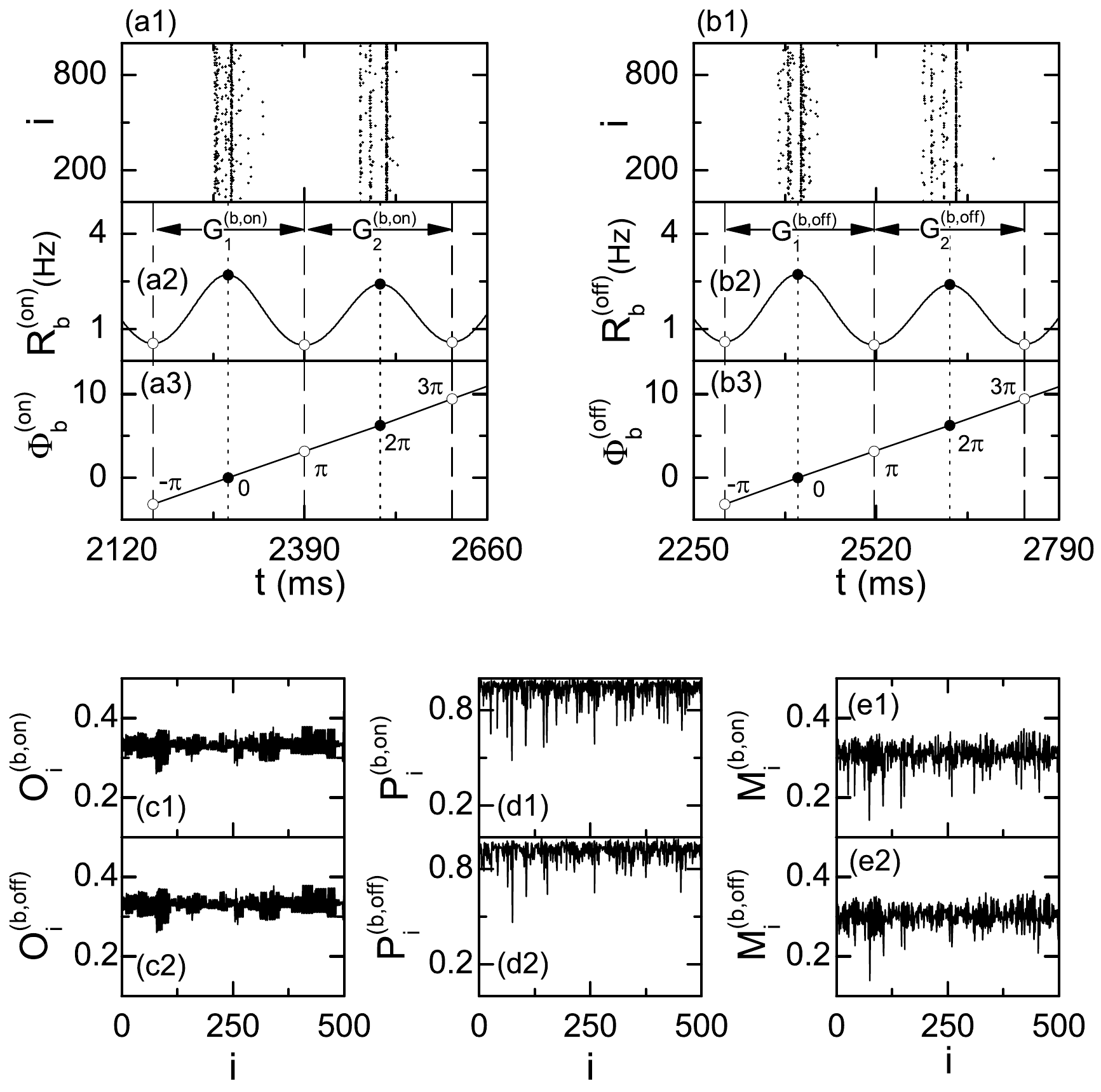}
\caption{Realistic statistical-mechanical bursting measures for measurement of the degree of burst synchronization, based on the IPBR kernel estimates, $R_b^{(on)}(t)$ and  $R_b^{(off)}(t)$, in an inhibitory population of $N$ $(=10^3$) globally-coupled bursting HR neurons for $I_{DC}=1.3$ and $J=0.3$ in the case of $D=0$. (a1) Raster plot of bursting onset times, (a2) time series of the IPBR kernel estimate $R_b^{(on)}(t)$, and (a3) the global bursting onset phase $\Phi_b^{(on)}(t)$. (b1) Raster plot of bursting offset times, (b2) time series of the IPBR kernel estimate $R_b^{(off)}(t)$, and (b3) the global bursting offset phase $\Phi_b^{(off)}(t)$. In (a2)-(a3) and (b2)-(b3), vertical dashed and dotted lines represent the times at which local minima and maxima (denoted by open and solid circles) of $R_b^{(on)}(t)$ and $R_b^{(off)}(t)$ occur, respectively, and $G_i^{(b,on)}$ [$G_i^{(b,off)}$] ($i=1,2$) denotes the $i$th global bursting onset (offset) cycle. Plots of (c1) [(c2)] $O_i^{(b,on)}$ [$O_i^{(b,off)}$] (occupation degree of bursting onset (offset) times in the $i$th global bursting onset (offset) cycle), (d1) [(d2)] $P_i^{(b,on)}$ [$P_i^{(b,off)}$] (pacing degree of bursting onset (offset) times in the $i$th global bursting onset (offset) cycle), and (e1) [(e2)] $M_i^{(b,on)}$ [$M_i^{(b,off)}$] [bursting onset (offset) measure in the $i$th global bursting onset (offset) cycle).
}
\label{fig:BM1}
\end{figure}

As a next step, within the burst-synchronized region ($0 \leq D < D^*_b$), we measure the degree of burst synchronization in terms of a realistic statistical-mechanical bursting measure $M_b$, based on the IPBR kernel estimates $R_b^{(on)}(t)$ and $R_b^{(off)}(t)$. Previously, a statistical-mechanical spiking measure, based on the ensemble-averaged global potential $X_G$, was developed for characterization of spike synchronization of spiking neurons \cite{Kim1}. However, the spiking measure, based on $X_G$, is practically inapplicable to the case of experimental neuroscience because to obtain $X_G$ in experiments is difficult. Hence, instead of $X_G$,  we used the experimentally-obtainable IPSR kernel estimate, and developed a refined version of statistical-mechanical spiking measure, based on the IPSR, to characterize spike synchronization of spiking neurons in both the experimental and the computational neuroscience \cite{Kim}. Here, we extend the realistic statistical-mechanical measure of spiking neurons (based on the IPSR) to the case of bursting neurons for measurement of the degree of the burst synchronization. As shown in Figs.~\ref{fig:Bursting}(e1)[(f1)]-\ref{fig:Bursting}(e5)[(f5)], burst synchronization may be well visualized in the raster plots of bursting onset (offset) times.  For the synchronous bursting case, bursting stripes (composed of bursting onset (offset) times and indicating population burst synchronization) appear in the raster plots. As an example, we consider a synchronous bursting case of $D=0$. The raster plot in Fig.~\ref{fig:BM1}(a1) is composed of partially-occupied and smeared stripes of bursting onset times. To measure the degree of  burst synchronization seen in the raster plot, we develop a statistical-mechanical bursting onset measure $M_b^{(on)}$, based on $R_b^{(on)}(t)$, by considering both the occupation pattern and the pacing pattern of the bursting onset times in the bursting onset stripes. The bursting onset measure $M_i^{(b,on)}$ of the $i$th bursting onset stripe is defined by the product of the occupation degree $O_i^{(b,on)}$ of bursting onset times (representing the density of the $i$th bursting onset stripe) and the pacing degree $P_i^{(b,on)}$ of bursting onset times (denoting the smearing of the $i$th bursting onset stripe):
\begin{equation}
M_i^{(b,on)} = O_i^{(b,on)} \cdot P_i^{(b,on)}.
\label{eq:BM1}
\end{equation}
The occupation degree $O_i^{(b,on)}$ of bursting onset times in the $i$th bursting onset stripe is given by the fraction of HR neurons which fire burstings:
\begin{equation}
   O_i^{(b,on)} = \frac {N_i^{(b,on)}} {N},
\label{eq:Occu}
\end{equation}
where $N_i^{(b,on)}$ is the number of HR neurons which fire burstings in the $i$th bursting onset stripe. For the full occupation $O_i^{(b,on)}=1$, while for the partial occupation $O_i^{(b,on)}<1$. The pacing degree $P_i^{(b,on)}$ of bursting onset times in the $i$th bursting onset stripe can be determined in a statistical-mechanical way by taking into account their contributions to the macroscopic IPBR kernel estimate $R_b^{(on)}(t)$. Figure \ref{fig:BM1}(a2) shows a time series of the IPBR kernel estimate $R_b^{(on)}(t)$ for $D=0$; local maxima and minima are denoted by solid and open circles, respectively. Obviously, central maxima of $R_b^{(on)}(t)$ between neighboring left and right minima of $R_b^{(on)}(t)$ coincide with centers of bursting onset stripes in the raster plot. The global bursting onset cycle starting from the left minimum of $R_b^{(on)}(t)$ which appears first after the transient time $(=2 \times 10^3$ ms) is regarded as the 1st one, which is denoted by $G_1^{(b,on)}$. The 2nd global bursting onset cycle $G_2^{(b,on)}$ begins from the next following right minimum of $G_1^{(b,on)}$, and so on. (The 1st global bursting onset cycle $G_1^{(b,on)}$ corresponds to the 2nd bursting onset stripe in Fig.~\ref{fig:Bursting}(e1) because the minimum of the global bursting onset cycle, corresponding to the 1st bursting onset stripe in Fig.~\ref{fig:Bursting}(e1), lies for $t<2 \times 10^3$ ms.) Then, we introduce an instantaneous global bursting onset phase $\Phi_b^{(on)}(t)$ of $R_b^{(on)}(t)$ via linear interpolation in the two successive subregions forming a global bursting onset cycle \cite{Kim,Kim1,GP}, as shown in Fig.~\ref{fig:BM1}(a3). The global bursting onset phase $\Phi_b^{(on)}(t)$ between the left minimum (corresponding to the beginning point of the $i$th global bursting onset cycle) and the central maximum is given by:
\begin{equation}
\Phi_b^{(on)}(t) = 2\pi(i-3/2) + \pi \left(
\frac{t-t_i^{(on,min)}}{t_i^{(on,max)}-t_i^{(on,min)}} \right)
 {\rm~~ for~} ~t_i^{(on,min)} \leq  t < t_i^{(on,max)}
~~(i=1,2,3,\dots),
\label{eq:Phi1}
\end{equation}
and $\Phi_b^{(on)}(t)$ between the central maximum and the right minimum (corresponding to the beginning point of the $(i+1)$th global bursting onset cycle) is given by
\begin{equation}
\Phi_b^{(on)}(t) = 2\pi(i-1) + \pi \left(
\frac{t-t_i^{(on,max)}}{t_{i+1}^{(on,min)}-t_i^{(on,max)}} \right)
 {\rm~~ for~} ~t_i^{(on,max)} \leq  t < t_{i+1}^{(on,min)}
~~(i=1,2,3,\dots),
\label{eq:Phi2}
\end{equation}
where $t_i^{(on,min)}$ is the beginning time of the $i$th global bursting onset cycle (i.e., the time at which the left minimum of $R_b^{(on)}(t)$ appears in the $i$th global bursting onset cycle) and $t_i^{(on,max)}$ is the time at which the maximum of $R_b^{(on)}(t)$ appears in the $i$th global bursting onset cycle. Then, the contribution of the $k$th microscopic bursting onset time in the $i$th bursting onset stripe occurring at the time $t_k^{(b,on)}$ to $R_b^{(on)}(t)$ is given by $\cos \Phi_k^{(b,on)}$, where $\Phi_k^{(b,on)}$ is the global bursting onset phase at the $k$th bursting onset time [i.e., $\Phi_k^{(b,on)} \equiv \Phi_b^{(on)}(t_k^{(b,on)})$]. A microscopic bursting onset time makes the most constructive (in-phase) contribution to $R_b^{(on)}(t)$ when the corresponding global onset phase $\Phi_k^{(b,on)}$ is $2 \pi n$ ($n=0,1,2, \dots$), while it makes the most destructive (anti-phase) contribution to $R_b^{(on)}(t)$ when $\Phi_k^{(b,on)}$ is $2 \pi (n-1/2)$. By averaging the contributions of all microscopic bursting onset times in the $i$th bursting onset stripe to $R_b^{(on)}(t)$, we obtain the pacing degree of bursting onset times in the $i$th bursting onset stripe:
\begin{equation}
 P_i^{(b,on)} ={ \frac {1} {B_i^{(on)}}} \sum_{k=1}^{B_i^{(on)}} \cos \Phi_k^{(b,on)},
\label{eq:Pacing}
\end{equation}
where $B_i^{(on)}$ is the total number of microscopic bursting onset times in the $i$th bursting onset stripe. By averaging $M_i^{(b,on)}$ of Eq.~(\ref{eq:BM1}) over a sufficiently large number $N_b^{(on)}$ of bursting onset stripes, we obtain the realistic statistical-mechanical bursting onset measure $M_b^{(on)}$, based on the IPSR kernel estimate $R_b^{(on)}(t)$:
\begin{equation}
M_b^{(on)} =  {\frac {1} {N_b}} \sum_{i=1}^{N_b^{(on)}} M_i^{(b,on)}.
\label{eq:BM2}
\end{equation}
For $D=0$ we follow $500$ bursting onset stripes and get $O_i^{(b,on)}$, $P_i^{(b,on)}$, and $M_i^{(b,on)}$ in each $i$th bursting onset stripe, which are shown in Figs.~\ref{fig:BM1}(c1), \ref{fig:BM1}(d1), and \ref{fig:BM1}(e1), respectively. Due to sparse burstings of individual HR neurons, the average occupation degree $O_b^{(on)}$ (=${\langle O_i^{(b,on)} \rangle}_b \simeq 0.33)$, where ${\langle \cdots \rangle}_b$ denotes the average over bursting onset stripes, is small. Hence, only a fraction (about 1/3) of the total HR neurons fire burstings in each bursting onset stripe. On the other hand, the average pacing degree  $P_b^{(on)}$ (=${\langle P_i^{(b,on)} \rangle}_b \simeq 0.94)$ is large in contrast to $O_b^{(on)}$. Consequently, the realistic statistical-mechanical bursting onset measure $M_b^{(on)}$ (=${\langle M_i^{(b,on)} \rangle}_b$), representing the degree of burst synchronization seen in the raster plot of bursting onset times, is about 0.31. The main reason for the low degree of burst synchronization is mainly due to partial occupation. In this way, the realistic statistical-mechanical bursting onset measure $M_b^{(on)}$ can be used effectively to measure the degree of burst synchronization because $M_b^{(on)}$ concerns not only the pacing degree, but also the occupation degree of bursting onset times in the bursting onset stripes of the raster plot.

\begin{figure}
\includegraphics[width=0.7\columnwidth]{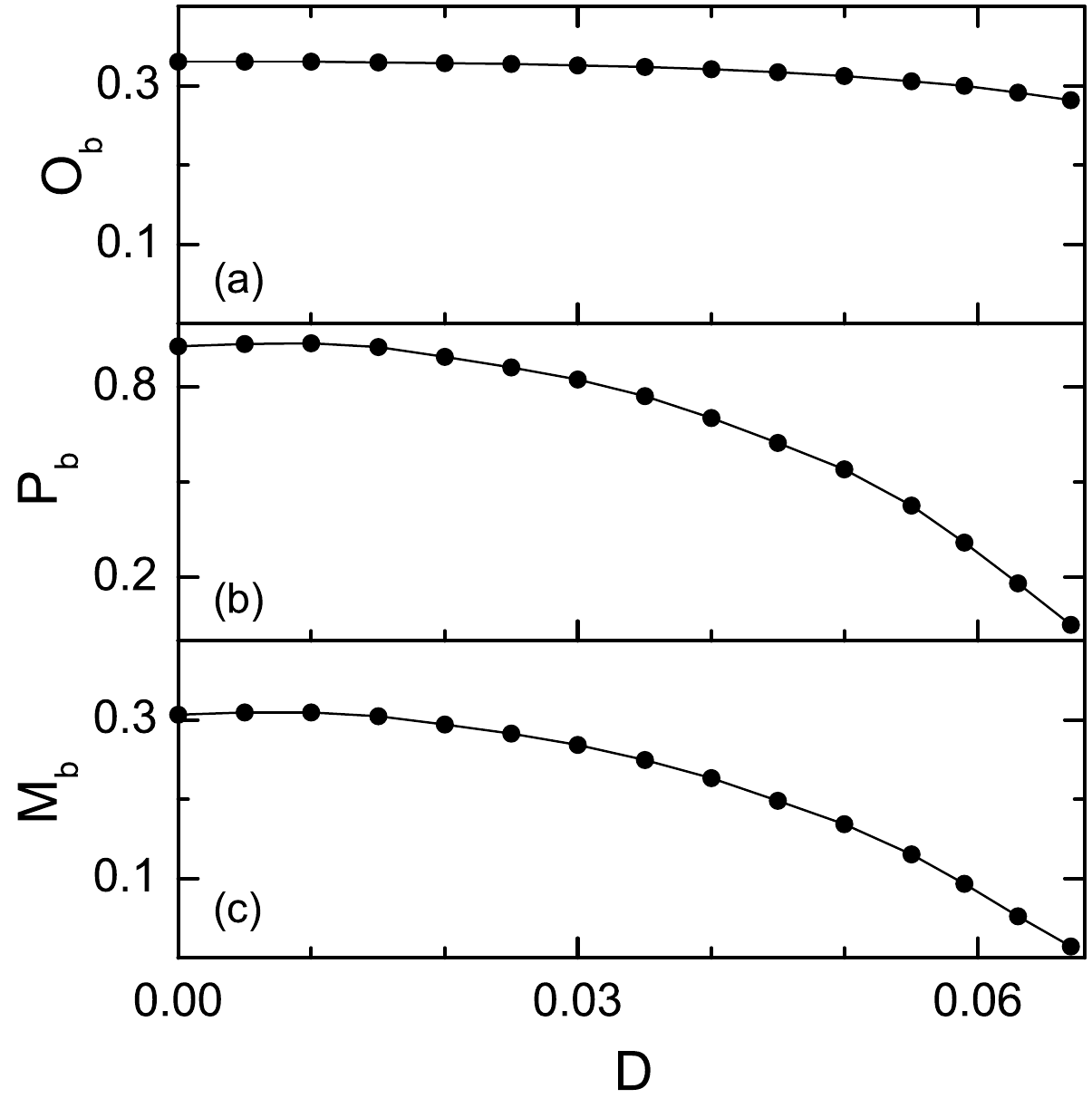}
\caption{Measurement of the degree of burst synchronization in terms of the realistic statistical-mechanical bursting measure $M_b$, based on the IPBR kernel estimates $R_b^{(on)}(t)$ and $R_b^{(off)}(t)$ in an inhibitory population of $N$ $(=10^3$) globally-coupled bursting HR neurons for $I_{DC}=1.3$ and $J=0.3$. (a) Plot of $O_b$ (average occupation degree of burstings) versus $D$. (b) Plot of $P_b$ (average pacing degree of burstings) versus $D$. (c) Plot of $M_b$ (realistic statistical-mechanical bursting measure) versus $D$. To obtain $O_b$, $P_b$, and $M_b$ in (a)-(c), we follow $500$ global bursting onset and 500 global bursting offset cycles for each $D$.
}
\label{fig:BM2}
\end{figure}

In addition to the above case of bursting onset times, we also consider population burst synchronization between the bursting offset times. Figures \ref{fig:BM1}(b1) and \ref{fig:BM1}(b2) show the raster plot composed of two stripes of bursting offset times and the corresponding IPBR $R_b^{(off)}$ for $D=0$, respectively; the 1st and 2nd global bursting offset cycles, $G_1^{(b,off)}$ and $G_2^{(b,off)}$, are shown. [Since the 1st global
cycle of offset times, $G_1^{(b,off)}$, follows the 1st global cycle of onset times, $G_1^{(b,on)}$, the 1st bursting offset stripe in Fig.~\ref{fig:BM1}(b1) corresponds to the 2nd bursting offset stripe in Fig.~\ref{fig:Bursting}(f1).] Then, as in the case of $\Phi_b^{(on)}(t)$,  one can introduce an instantaneous global bursting offset phase $\Phi_b^{(off)}(t)$ of $R_b^{(off)}(t)$ via linear interpolation in the two successive subregions forming a global bursting offset cycle [see Fig.~\ref{fig:BM1}(b3)]. Similar to the case of bursting onset times, we measure the degree of burst synchronization seen in the raster plot of bursting offset times in terms of a statistical-mechanical bursting offset measure $M_b^{(off)}$, based on $R_b^{(off)}(t)$, by considering both the occupation pattern and the pacing pattern of the bursting offset times in the bursting offset stripes. The bursting offset measure $M_i^{(b,off)}$ in the $i$th bursting offset stripe also is defined by the product of the occupation degree $O_i^{(b,off)}$ of bursting offset times and the pacing degree $P_i^{(b,off)}$ of bursting offset times in the $i$th bursting offset stripe. We also follow $500$ bursting offset stripes and get $O_i^{(b,off)}$, $P_i^{(b,off)}$, and $M_i^{(b,off)}$ in each $i$th bursting offset stripe for $D=0$, which are shown in Figs.~\ref{fig:BM1}(c2), \ref{fig:BM1}(d2), and \ref{fig:BM1}(e2), respectively. For this case of bursting offset times, $O_b^{(off)}$ (=${\langle O_i^{(b,off)} \rangle}_b) \simeq 0.33$, $P_b^{(off)}$ (=${\langle P_i^{(b,off)} \rangle}_b) \simeq 0.92$, and $M_b^{(off)}$ (=${\langle M_i^{(b,off)} \rangle}_b$) $\simeq 0.30$. The pacing degree of offset times is a little smaller than the pacing degree of the onset times ($P_b^{(on)} \simeq 0.94$), although the occupation degrees $(\simeq 0.33)$ of the onset and offset times are the same. We take into consideration both cases of the onset and offset times equally and define the average occupation degree $O_b$, the average pacing degree $P_b$, and the statistical-mechanical bursting measure $M_b$ as follows:
\begin{equation}
 O_b = [O_b^{(on)} + O_b^{(off)}]/2,~~P_b = [P_b^{(on)} + P_b^{(off)}]/2,~~{\rm{and}}~~M_b = [M_b^{(on)} + M_b^{(off)}]/2.
\label{eq:BM3}
\end{equation}
By increasing the noise intensity $D$, we follow 500 bursting onset and 500 bursting offset stripes and characterize burst synchronization in terms of $O_b$ (average occupation degree), $P_b$ (average pacing degree), and $M_b$ (statistical-mechanical bursting measure) for 15 values of $D$ in the whole region of burst synchronization $[0 \leq D < D^*_b (\simeq 0.068)$], and the results are shown in Figs.~\ref{fig:BM2}(a)-\ref{fig:BM2}(c). As $D$ is increased, the average occupation degree $O_b$ (denoting the average density of bursting stripes in the raster plot) decreases very slowly around $O_b \sim 0.32$, because a little tendency for noise-induced skipping of burstings in individual HR neurons occurs \cite{Order3}. On the other hand, with increasing $D$, the average pacing degree $P_b$ (representing the average smearing of the bursting stripes in the raster plot) decreases rapidly due to destructive role of noise spoiling burst synchronization. The statistical-mechanical bursting measure $M_b$ also makes a rapid decrease because of a rapid drop in $P_b$. Both $P_b$ and $M_b$ show quadratic decreases because they are well fitted with quadratic functions: $P_b \simeq -254.18\,D^2 + 4.35\,D+0.93$ and $M_b \simeq -73.26\,D^2 + 1.26\,D +0.31$. In this way, we measure the degree of burst synchronization in terms of the realistic statistical-mechanical bursting measure $M_b$ in the whole synchronized region, and find that $M_b$ reflects the degree of burst synchronization seen in the raster plot of onset and offset times very well.

Unlike the case of spiking neurons (showing only the spike synchronization), bursting neurons may exhibit both the burst and the spike synchronizations. From now on, we investigate the intraburst spike synchronization of bursting HR neurons by varying the noise intensity $D$.
Figures \ref{fig:Spiking}(a1)-\ref{fig:Spiking}(a6) and Figures \ref{fig:Spiking}(b1)-\ref{fig:Spiking}(b6) show the raster plots of intraburst spikes and the corresponding IPFR kernel estimates $R(t)$ during the 1st global bursting cycle of the low-pass filtered IPBR $R_b(t)$, respectively for various values of $D$: synchronized spiking states for $D=0$, 0.005, 0.01, and 0.02, and unsynchronized spiking states for $D=0.04$ and 0.08. As mentioned above, $R(t)$ exhibits the whole combined population behaviors including the burst and spike synchronizations with both the slow bursting and the fast spiking timescales. Hence, through band-pass filtering of $R(t)$ [with the lower and the higher cut-off frequencies of 30 Hz (high-pass filter) and 90 Hz (low-pass filer)], we obtain the IPSRs $R_s(t)$, which are shown in Figs.~\ref{fig:Spiking}(c1)-\ref{fig:Spiking}(c6). Then, the intraburst spike synchronization may be well described in terms of the IPSR $R_s(t)$. For $D=0$, clear 8 ``spiking stripes'' (composed of spikes and indicating population spike synchronization) appear in the bursting band of the 1st global bursting cycle of the IPBR $R_b(t)$ [see Fig.~\ref{fig:Spiking}(a1)], and the IPFR kernel estimate $R(t)$ exhibits a bursting activity [i.e., fast spikes appear on a slow wave in $R(t)$] due to the complete synchronization (including both the burst and spike synchronizations), as shown in Fig.~\ref{fig:Spiking}(b1). However, the band-pass filtered IPSR $R_s(t)$ shows only the fast spiking oscillations (without a slow wave) with the population spiking frequency $f_s$ $(\simeq 68.5$ Hz) [see Fig.~\ref{fig:Spiking}(c1)]. As $D$ is increased, spiking stripes in the bursting band become more and more smeared (e.g., see the cases of $D=0.005$, 0.01, and 0.02). As a result, the amplitude of the IPSR $R_s(t)$ decreases due to loss of spike synchronization. Eventually, when passing the spiking noise threshold $D^*_s$ $(\simeq 0.032)$, spikes become completely scattered within the bursting band (i.e., intraburst spikes become completely incoherent), and hence complete loss of spike synchronization occurs in the bursting band. As an example, see the case of $D=0.04$. For this case, the IPSR $R_s(t)$ becomes nearly stationary, while the IPFR kernel estimate $R(t)$ shows a slow-wave oscillation (without spikes) due to the burst synchronization. Thus, for $D>D^*_s$ only the burst synchronization may occur. With further increase in $D$, the incoherent bursting band expands, fills the whole region of the global bursting cycle, and overlaps with nearest bursting bands. Consequently, complete loss of burst synchronization also occurs when passing the larger bursting noise threshold $D^*_b$ $(\simeq 0.068)$. Thus, for $D>D^*_b$ completely unsynchronized states with nearly stationary $R(t)$ appear (e.g., see the case of $D=0.08$).

For characterization of the intraburst spiking transition, we employ the experimentally-obtainable IPSR $R_s(t)$ (which is obtained from the IPFR kernel estimate $R(t)$ via band-pass filtering), and develop a realistic spiking order parameter ${\cal{O}}_s$, which may be applicable in both the computational and the experimental neuroscience. The mean square deviation of $R_s(t)$ in the $i$th global bursting cycle,
\begin{equation}
{\cal{O}}_s^{(i)} \equiv \overline{(R_s(t) - \overline{R_s(t)})^2},
\end{equation}
plays the role of a spiking order parameter ${\cal{O}}_s^{(i)}$ in the $i$th global bursting cycle of the low-pass filtered IPBR $R_b(t)$.
By averaging ${\cal{O}}_s^{(i)}$ over a sufficiently large number $N_b$ of global bursting cycles, we obtain the realistic spiking order parameter:
\begin{equation}
{\cal{O}}_s =  {\frac {1} {N_b}} \sum_{i=1}^{N_b} {\cal{O}}_s^{(i)}.
\end{equation}
By following $500$ bursting cycles, we obtain the spiking order parameter ${\cal{O}}_s$. Figure \ref{fig:Spiking}(d) shows plots of ${\cal{O}}_s$ versus $D$. For $D < D^*_s$ $(\simeq 0.032$), synchronized spiking states exist because the values of ${\cal {O}}_s$ become saturated to non-zero limit values in the thermodynamic limit of $N \rightarrow \infty$. However, when passing the spiking noise threshold $D^*_s$, the spiking order parameter ${\cal{O}}_s$ tends to zero as $N \rightarrow \infty$, and hence a transition to unsynchronized spiking states occurs because the noise spoils the spike synchronization. In this way, the spiking noise threshold $D^*_s$ may be well determined through calculation of the realistic spiking order parameter ${\cal{O}}_s$.

\begin{figure}
\includegraphics[width=0.6\columnwidth]{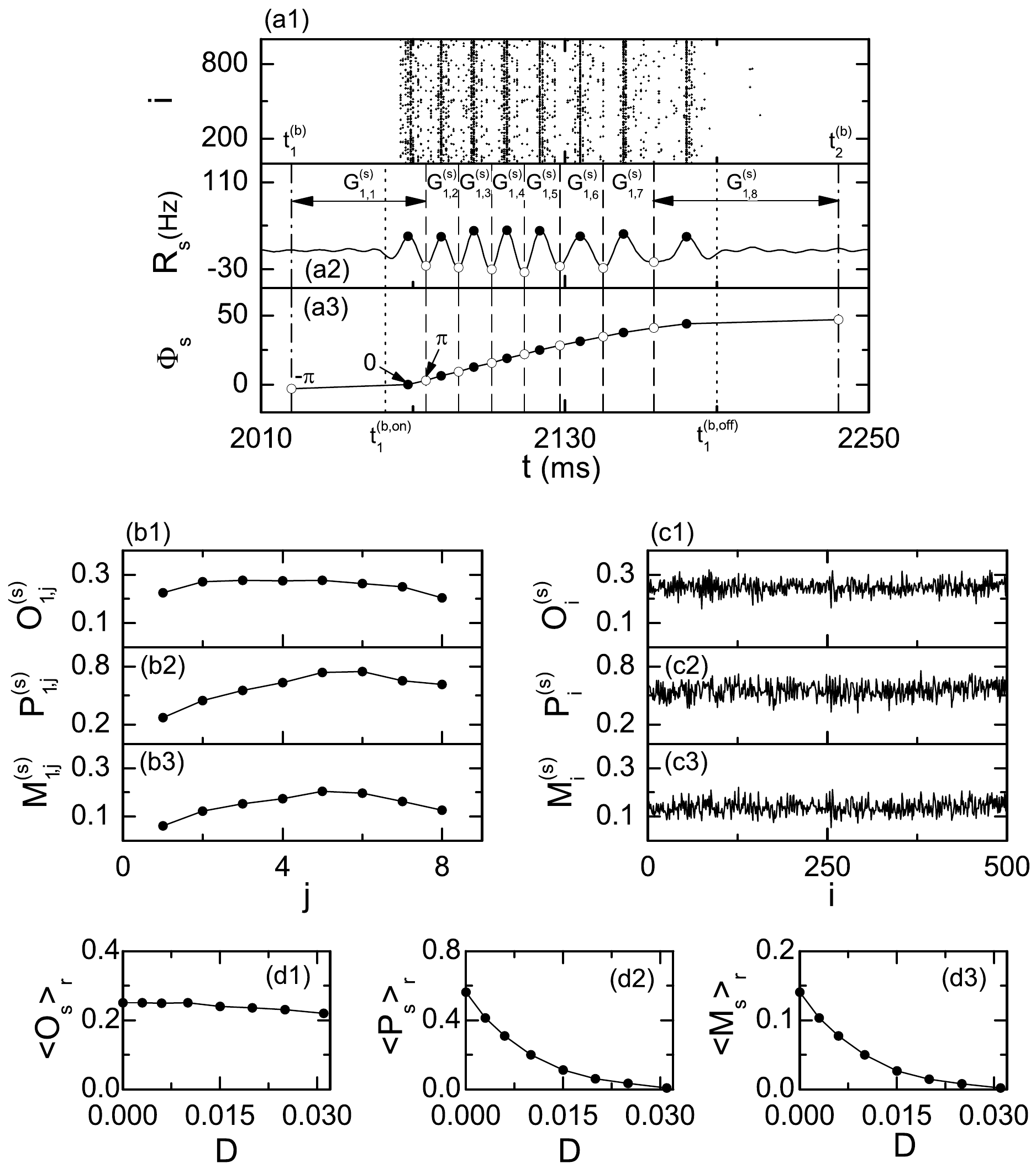}
\linespread{1.2}
\caption{Statistical-mechanical intraburst spiking measure for measurement of the degree of intraburst spike synchronization, based on the IPSR kernel estimates, $R_s(t)$ in an inhibitory population of $N$ $(=10^3$) globally-coupled bursting HR neurons for $I_{DC}=1.3$ and $J=0.3$. For $D=0$, (a1) a magnified raster plot of neural spikes, (a2) time series of the IPSR $R_s(t)$, and (a3) time series of the global spiking phase $\Phi_s(t)$ in the 1st global bursting cycle of $R_b(t)$ [bounded by vertical dash-dotted lines: $t_1^{(b)} (=2022 {\rm ms}) < t < t_2^{(b)} (=2238 {\rm ms})$]. Within the 1st global bursting cycle, the bursting band [bounded by vertical dotted lines: $t_1^{(b,on)} (=2059 {\rm ms}) < t < t_2^{(b,off)} (=2190 {\rm ms})$] in (a1), corresponding to the 1st global active phase, is composed of 8 stripes of spikes; $t_1^{(b,on)}$ (maximum of $R_b^{(on)}(t)$ within the 1st global bursting cycle) is the global bursting onset time, and $t_1^{(b,off)}$ (maximum of $R_b^{(off)}(t)$ within the 1st global bursting cycle) is the global bursting offset time. In the bursting band, the maxima (minima) of $R_s(t)$ are denoted by solid (open) circles, and 8 spiking cycles $G_{1,j}^{(s)}$ $(j=1, ... , 8)$ exist in the 1st global bursting cycle. For $D=0$, (b1) plot of $O_{1,j}^{(s)}$ (occupation degree of  spikes), (b2) plot of $P_{1,j}^{(s)}$ (pacing degree of spikes), and (b3) $M_{1,j}^{(s)}$ (spiking measure) in the $j$th spiking phase $G_{1,j}^{(s)}$ of the 1st global bursting cycle of $R_b(t)$ versus $j$. For $D=0$, (c1) plot of $O_i^{(s)}$ (occupation degree of spikes), (c2) plot of $P_i^{(s)}$ (pacing degree of spikes), and (c3) $M_i^{(s)}$ (spiking measure) in the $i$th global bursting cycle versus $i$. Measurement of the degree of intraburst spike synchronization: (d1) plot of ${\langle O_s \rangle}_r$ (average occupation degree of spikes). (d2) plot of ${\langle P_s \rangle}_r$ (average pacing degree of spikes), and (d3) plot of ${\langle M_s \rangle}_r$ (average statistical-mechanical intraburst spiking measure) versus $D$. For each $D$, we follow 100 bursting cycles in each realization, and obtain ${\langle O_s \rangle}_r$, ${\langle P_s \rangle}_r$, and ${\langle M_s \rangle}_r$ via average over 20 realizations.
}
\label{fig:SM}
\end{figure}

Finally, within the whole region of the intraburst spike synchronization ($0 \leq  D < D^*_s$), we measure the degree of intraburst spike synchronization in terms of a realistic statistical-mechanical spiking measure $M_s$, based on the IPSR $R_s(t)$. As shown in Figs.~\ref{fig:Spiking}(a1)-\ref{fig:Spiking}(a6), spike synchronization may be well visualized in the raster plot of spikes. For the synchronous spiking case, spiking stripes (composed of spikes and indicating population spike synchronization) appear within the bursting bands of the raster plot. As an example, we consider a synchronous spiking case of $D=0$. Figures \ref{fig:SM}(a1) and \ref{fig:SM}(a2) show a magnified raster plot of neural spikes and the IPSR $R_s(t)$, corresponding to the 1st global bursting cycle of the low-pass filtered IPBR $R_b(t)$ [bounded by a vertical dash-dotted lines: $t_1^{(b)} (=2022 {\rm ms}) < t < t_2^{(b)} (=2238 {\rm ms})$]. Within the 1st global cycle, the bursting band [bounded by the vertical dotted lines: $t_1^{(b,on)} (=2059 {\rm ms}) < t < t_2^{(b,off)} (=2190 {\rm ms})$], corresponding to the 1st global active phase, is composed of 8 stripes of spikes, as shown in Fig.~\ref{fig:SM}(a1); $t_1^{(b,on)}$ (maximum of $R_b^{(on)}(t)$ within the 1st global bursting cycle) is the global active phase onset time, and $t_1^{(b,off)}$ (maximum of $R_b^{(off)}(t)$ within the 1st global bursting cycle) is the global active phase offset time. In the bursting band, the maxima (minima) of $R_s(t)$ are denoted by solid (open) circles, and 8 global spiking cycles $G_{1,j}^{(s)}$ $(j=1, ... , 8)$ exist in the 1st global bursting cycle of $R_b(t)$, as shown in Fig.~\ref{fig:SM}(a2). For $1<j<8$, each $j$th global spiking cycle $G_{1,j}^{(s)}$, containing the $j$th maximum of $R_s(t)$, begins at the left nearest-neighboring minimum of $R_s(t)$ and ends at the right nearest-neighboring minimum of $R_s(t)$, while for both extreme cases of $j=1$ and 8, $G_{1,1}^{(s)}$ begins at $t_1^{(b)}$ [the beginning time of the 1st global bursting cycle of $R_b(t)$] and $G_{1,8}^{(s)}$ ends at $t_2^{(b)}$ [the ending time of the 1st global bursting cycle of $R_b(t)$]. Then, as in the case of the global bursting phase, one can introduce an instantaneous global spiking phase $\Phi_s(t)$ of $R_s(t)$ via linear interpolation in the two successive subregions (the left subregion joining the left beginning point and the central maximum and the right subregion joining the central maximum and the right ending point) forming a global spiking cycle [see Fig.~\ref{fig:SM}(a3)]. Similar to the case of burst synchronization, we measure the degree of the intraburst spike synchronization seen in the raster plot in terms of a statistical-mechanical spiking measure, based on $R_s(t)$, by considering both the occupation pattern and the pacing pattern of spikes in the global spiking cycles. The spiking measure $M_{1,j}^{(s)}$ of the $j$th global spiking cycle in the 1st global bursting cycle is defined by the product of the occupation degree $O_{1,j}^{(s)}$ of spikes (representing the density of spikes in the $j$th global spiking cycle) and the pacing degree $P_{1,j}^{(s)}$ of spikes (denoting the smearing of spikes in the $j$th global spiking cycle). Figures \ref{fig:SM}(b1)-\ref{fig:SM}(b3) show the plots of $O_{1,j}^{(s)}$, $P_{1,j}^{(s)}$, and $M_{1,j}^{(s)}$, respectively. For the 1st global bursting cycle, the spiking-averaged occupation degree $O_1^{(s)}$ (=${\langle O_{1,j}^{(s)} \rangle}_s$) $ \simeq 0.25$, the spiking-averaged pacing degree $P_1^{(s)}$ (=${\langle P_{1,j}^{(s)} \rangle}_s$) $ \simeq 0.56$, and the spiking-averaged statistical-mechanical spiking measure $M_1^{(s)}$ (=${\langle M_{1,j}^{(s)} \rangle}_b$) $ \simeq 0.14$, where ${\langle \cdots \rangle}_s$ represents the average over the spiking cycles. We also follow $500$ bursting cycles and get $O_i^{(s)}$, $P_i^{(s)}$, and $M_i^{(s)}$ in each $i$th global bursting cycle for $D=0$, which are shown in Figs.~\ref{fig:SM}(c1), \ref{fig:SM}(c2), and \ref{fig:SM}(c3), respectively. Then, through average over all bursting cycles, we obtain the bursting-averaged occupation degree $O_s$ (=${\langle O_i^{(s)} \rangle}_b \simeq 0.25)$, the bursting-averaged pacing degree $P_s$ (=${\langle P_i^{(s)} \rangle}_b \simeq 0.56)$, and the bursting-averaged statistical-mechanical spiking measure $M_s$ (=${\langle M_i^{(s)} \rangle}_b \simeq 0.14)$ for $D=0$. We note that $O_s$, $P_s$, and $M_s$ are obtained through double-averaging ${\langle {\langle \cdots \rangle}_s \rangle}_b$ over the spiking and bursting cycles. When compared with the bursting case of $O_b \simeq 0.33$ and $P_b \simeq 0.93$ for $D=0$, a fraction (about 3/4) of the HR neurons exhibiting the bursting active phases fire spikings in the spiking cycles, and the pacing degree of spikes $P_s$ is about 60 percentage of the pacing degree of burstings $P_b$. Consequently, the statistical-mechanical spiking measure $M_s$ becomes about 45 percentage of the statistical-mechanical bursting measure $M_b$ for $D=0$. We increase the noise intensity $D$ and obtain $O_s$, $P_s$, and $M_s$. However, as will be seen below, with increasing $D$, $P_s$ decreases very rapidly in an exponential way, in contrast to the bursting case. Hence, for more accurate results, we repeat the process to get $O_s$, $P_s$, and $M_s$ for multiple realizations. Thus, we obtain ${\langle O_s \rangle}_r$ (average occupation degree of spikes in the global spiking cycles), ${\langle P_s \rangle}_r$ (average pacing degree of spikes in the global spiking cycles), and ${\langle M_s \rangle}_r$ (average statistical-mechanical spiking measure in the global spiking cycles) through average over all realizations. For each realization, we follow 100 bursting cycles, and obtain ${\langle O_s \rangle}_r$, ${\langle P_s \rangle}_r$, and ${\langle M_s \rangle}_r$ via average over 20 realizations. Through these multiple-realization simulations, we characterize intraburst spike synchronization in terms of ${\langle O_s \rangle}_r$, ${\langle P_s \rangle}_r$, and ${\langle M_s \rangle}_r$ for 8 values of $D$ in the whole region of spike synchronization [$0 \leq D < D^*_s (\simeq 0.032)$], which are shown in Figs.~\ref{fig:SM}(d1)-\ref{fig:SM}(d3), respectively. As $D$ is increased, the average occupation degree ${\langle O_s \rangle}_r$ decreases very slowly around ${\langle O_s \rangle}_r \sim 0.24$ due to a little tendency for noise-induced subtracting of spikes in individual HR neurons, while the average pacing degree ${\langle P_s \rangle}_r$ decreases very rapidly due to destructive role of noise spoiling spike synchronization. The average statistical-mechanical spiking measure ${\langle M_s \rangle}_r$ also makes a rapid decrease because of a rapid drop in ${\langle P_s \rangle}_r$. Both ${\langle P_s \rangle}_r$ and ${\langle M_s \rangle}_r$ exhibit exponential decreases because they are well fitted with exponential functions: ${\langle P_s \rangle}_r \simeq 0.58~e^{-97.55\,D}-0.019$ and ${\langle M_s \rangle}_r \simeq 0.15~e^{-98.05\,D}-0.005$. In this way, we measure the degree of intraburst spike synchronization in terms of the realistic statistical-mechanical spiking measure ${\langle M_s \rangle}_r$ in the whole synchronized region, and find that ${\langle M_s \rangle}_r$ reflects the degree of intraburst spike synchronization seen in the raster plot very well. Finally, we note that the exponential loss in the degree of spike synchronization is much faster than the quadratic loss in the degree of the burst synchronization. As a result, the break-up of the spike synchronization occurs first at the smaller spiking noise threshold $D^*_s$ $(\simeq 0.032)$, and then the burst synchronization disappears at the larger bursting noise threshold  $D^*_b$ $(\simeq 0.068)$.

\section{Summary} \label{sec:SUM}
We have extended the order parameter and the statistical-mechanical measure to the case of bursting neurons. Their usefulness for characterization of the burst and spike synchronizations has been shown in explicit examples of bursting HR neurons by varying the noise intensity $D$. We note that population synchronization may be well visualized in the raster plot of neural spikes which may be obtained in experiments. Unlike the case of spiking neurons, bursting neurons show firing patterns with two timescales: a fast spiking timescale and a slow bursting timescale that modulates the spiking activity. Hence, the IPFR kernel estimate $R(t)$, which is obtained from the raster plot of spikes, shows collective behaviors with both the slow bursting and the fast spiking timescales. For our purpose, we separate the slow bursting and the fast spiking timescales via frequency filtering, and  decompose the IPFR kernel estimate $R(t)$ into the IPBR $R_b(t)$ and the IPSR $R_s(t)$. Based on $R_b(t)$ and $R_s(t)$, we have developed the bursting and spiking order parameters ${\cal {O}}_b$ and ${\cal {O}}_s$ which may be used to determine the bursting and spiking noise thresholds, $D^*_b$ and $D^*_s$, for the burst and spike synchronizations. When passing $D^*_b$ and $D^*_s$, loss of the burst and spike synchronizations occurs due to destructive role of noise spoiling the burst and spike synchronizations, respectively. As a next step, the degree of burst synchronization seen in the raster plot of bursting onset or offset times has been well measured in the whole region of burst synchronization in terms of a statistical-mechanical bursting measure $M_b$, introduced by considering both the occupation and the pacing patterns of bursting  onset or offset times in the raster plot. Similarly, we have also developed a statistical-mechanical spiking measure $M_s$, based on $R_s(t)$, and measured the degree of the intraburst spike synchronization well. Thus, the statistical-mechanical bursting and spiking measures have been found to reflect both the occupation and the pacing degrees of bursting onset or offset times and spikes seen in the raster plot very well. Furthermore, it has also been found that the exponential loss in the degree of spike synchronization is much faster than the quadratic loss in the degree of the burst synchronization. Hence, the intraburst spike synchronization breaks up first at the smaller spiking noise threshold $D^*_s$ $(\simeq 0.032)$, and then the burst synchronization disappears at the larger bursting noise threshold  $D^*_b$ $(\simeq 0.068)$. Consequently, we have shown in explicit examples that the order parameters and the statistical-mechanical measures may be effectively used to determine the bursting and spiking thresholds for the burst and the spike synchronizations and also to quantitatively measure the degree of the burst and the spike synchronizations, respectively.

\begin{acknowledgments}
This research was supported by Basic Science Research Program through the National Research Foundation of Korea (NRF) funded by the Ministry of Education (Grant No. 2013057789).
\end{acknowledgments}

\newpage

\newpage

\newpage

\newpage

\newpage

\newpage

\newpage

\end{document}